# ELECTRON WAVE FUNCTION IN THE FIELD OF ONE DIMENSIONAL IRREGULAR LAYERED STRUCTURE

## A.Zh. Khachatrian

*State Engineering University of Armenia, 375049,Terian 109, Yerevan, Armenia*

[akhachat@www.physdep.r.am](akhachat@www.physdep.r.am)

A method is proposed to find the wave function of an electron moving infinitely in the field of an arbitrary 1D layer structure with two different homogeneous semi-infinite boundaries. It is shown that in general the problem reduces to solution of a set of two linear difference equations. The proposed approach is discussed on a base of two cases: a structure of periodically placed identical rectangular potentials and a non-ordered structure with certain distortion of periodicity and potential identity.

## 1. Introduction

It is known that the determination of electron wave function and spectrum of electron's bound states in a field of an arbitrary potential has both physical [1] and practical [2, 3] importance. In general, the analytic approach to this class of problems for 2- and 3D systems, when the electron's potential energy is an arbitrary or randomly changing from point to point function, is related with significant mathematical difficulties, hence they are mainly considered numerically.

In parallel, 1D models are being intensively explored during many years. Those are still actual and are under investigation intensively up to date [4 - 11]. This is due to several reasons. First, 1 D models are of physical interest *per se* and actual for many applied problems. Second, understanding of the 1D phenomenon provides opportunities for better understanding of 2- and 3D cases. Third, the development of general solution methods for 1D cases may serve as a base for development of powerful methods for multidimensional cases.

Let us consider electron motion in a field of 1D potential

$$\left(-\frac{d^2}{dx^2} + u(x)\right)\psi(x) = \varepsilon \psi(x), \tag{1}$$

where $u(x) = \hbar^2/2mU(x)$, $\varepsilon = 2mE/\hbar^2$, and $U(x)$, $E$ are electron's potential and full energies respectively. Let us assume that $u(x)$ has a form

$$U(x) = \begin{cases} V_1 = const, & when \quad x \to -\infty, \\ V_2 = const, & when \quad x \to +\infty, \end{cases}$$

and $V_1$, $V_2$ in general have different values. Further we will take the minimum potential value as the original point of energy, so that at $V_1 = V_2 = 0$ the spectrum of electron states will be continuous.

It is known that three following cases are considered depending on wave function behavior at $x \to -\infty$ and $x \to +\infty$. The first one is the case of electron's infinite motion, when the wave function is non-zero throughout the whole space. Such a wave function corresponds, e.g., to an electron with energy $e > V_1, V_2$ incidenting from the second semi-infinite medium and partially penetrates in the first semi-infinite medium. The second case corresponds to a wave function that is non-zero only in the semi-infinite region of the 1D space. Such a wave function (for being definite, let it be $V_1 > V_2$) corresponds to an electron with energy $e < V_1$ and $e > V_2$ incidenting from the second semi-infinite medium and, being fully reflected, again propagates in the second medium. The condition $e < V_1$ provides wave damping in the first medium. The last third case corresponds to an electron which wave function notably differs from zero only in the layer region and decreases while deepening into the first and second semi-infinite media. This wave function describes the so-called bound state in which the electron with energy $e < V_1, V_2$ might occur. It is important that for the first two cases the energy spectrum of states is continuous, while it is discrete in the third case.

The problem of wave function determination for the listed three cases and electron spectrum for bound states has been considered by numerous precise and approximate methods, of which most common are Green functions, transfer matrices, quasi-classical approximation, perturbation theory, et al. [1, 12 - 16].

## 2. Wave Function of an Electron Scattered in a Field of Irregular Structure of Rectangular Potentials

Let us consider the problem of the wave function determination for an electron moving stationary and infinitely in the potential field of a 1D irregular lattice structure consisting of randomly placed homogeneous layers of different thickness. For more generality we assume that the non-ordered structure has left and right boundaries with two different homogeneous semi-infinite media with potentials $V_1$ and $V_2$ respectively, so that the potential throughout the whole space can be present in the following form:

$$V(x) = \begin{cases} V_1 = const, & x \leq 0, \\ \sum_{n=1}^{N} U_n q(x - x_n + d_n/2) q(x_n + d_n/2 - x), & 0 < x < d, \\ V_2 = const, & x \geq d \end{cases} \quad (2)$$

According to (2), in the interval $0 < x < d$ the potential is a layered structure consisting of N rectangular potentials (2), where $U_n$ and $d_n$ are the value and width of n-th rectangular potential, $x_n$ is the coordinate of its medium point. Note that in (2) we assume $x_1 - d_1/2 \geq 0$, $x_n + d_n/2 \leq d$, $x_n + d_n/2 \leq x_{n+1} - d_{n+1}/2$ ($n = 1, 2, \cdots, N-1$).

While considering electron's infinite motion in a 1D field, it is common to restrict with determination of wave function's asymptotes at infinities. Thus, in case of electron scattering on a 1D potential, only electron reflection and penetration amplitudes are determined, while the wave function within the potential remains undetermined. However, in many cases of physical interest it is necessary to know the delocalized wave function not only in asymptotes, but in the whole space as well. Such a necessity arises, e.g., at investigation of optical absorption characteristics in semiconductor heterostructures, related to electron transit from bound state of discrete spectrum into a delocalized state of the continuous part of spectrum [17, 18].

Let us introduce following denotes:

$$k_{10} = \sqrt{e - V_1}, \quad k_{02} = \sqrt{e - V_2}, \quad k_0 = \sqrt{e} \text{ and } k_n = \sqrt{e - U_n} \ (n = 1, 2 \cdots N). \quad (3)$$

Let the initial electron wave with unit amplitude incide on an unordered structure from the first semi-infinite medium. Then the solution of (1) with potential (2) in the whole space is wave function that can be represented in a following form:

$$\mathbf{y}(x) = \begin{cases} \exp\{ik_{10}x\} + R_{1,2}^N \exp\{-ik_{10}x\}, & x < 0 \\ a_1 \exp\{ik_0 x\} + b_1 \exp\{-ik_0 x\}, & 0 < x < x_1 - d_1/2 \\ c_1 \exp\{ik_1 x\} + d_1 \exp\{-ik_1 x\}, & x_1 - d_1/2 < x < x_1 + d_1/2 \\ \dots \dots \dots \dots \dots \dots \\ c_N \exp\{ik_N x\} + d_N \exp\{-ik_N x\}, & x_N - d_N/2 < x < x_N + d_N/2 \\ a_{N+1} \exp\{ik_0 x\} + b_{N+1} \exp\{-ik_0 x\}, & x_N + d_N/2 < x < d \\ T_{1,2}^N \exp\{ik_{02} x\}, & x > d \end{cases} \quad (4)$$

where $R_{1,2}^N$ and $T_{1,2}^N$ are electron's reflection and transit for potential (2).

To be a wave function, (4) and its derivatives must be continuous functions in points $x = 0, \dots, x_j - d_j/2, x_j + d_j/2, \dots, d$, which is equivalent to the condition of probability flow conservation. The continuity of $\mathbf{y}(x)$ and $d\mathbf{y}(x)/dx$ in all points of the space can be achieved by a proper choice of coefficients $R_{1,2}^N, T_{1,2}^N, c_n, d_n$ $(n = 1, 2, \dots, N)$ and $a_n, b_n$ $(n = 1, 2, \dots, N+1)$. One can easily see that the continuity requirement is equivalent to the fact that the coefficients in solution (4) should satisfy to a set of $4(N+1)$ linear homogeneous equations. In general, the solution of this set is a difficult mathematical problem. However, below we will show that the problem of determination of wave function coefficients (or the wave function in the whole space) in general be reduced to a solution of a set of two linear first order difference equations. Prior proceeding to solution of the stated problem, let us introduce some useful denotes:

$$t_{1,0} = \frac{2k_{10}}{k_1 + k_0}, \quad r_{1,0} = \frac{k_{10} - k_0}{k_{10} + k_0}, \tag{5}$$

$$t_{0,2} = \frac{2k_0}{k_0 + k_{02}} \exp\{i(k_0 - k_{02})d\}, \quad r_{0,2} = \frac{k_0 - k_{02}}{k_0 + k_{02}} \exp\{i2k_0 d\}, \tag{6}$$

$$t_{0,n} = \frac{2k_0}{k_0 + k_n} \exp\{i(k_0 - k_n)(x_n - d_n/2)\}, \quad r_{0,n} = \frac{k_0 - k_n}{k_0 + k_n} \exp\{i2k_0(x_n - d_n/2)\}, \tag{7}$$

$$\frac{\exp\{-ik_0 d_n\}}{t_n} = \left\{\cos k_n d_n - i\frac{k_n^2 + k_0^2}{2k_n k_0}\sin k_n d_n\right\}, \quad \frac{r_n}{t_n} = i\exp\{ik_0 x_n\}\frac{k_0^2 - k_n^2}{2k_n k_0}\sin k_n d_n. \tag{8}$$

Note that in (7) and (8) $n$ is considered as a variable $(n = 1, 2, ..., N)$. According to denotes (5), (6), $t_{1,0}$, $r_{1,0}$ ($t_{0,2}$, $r_{0,2}$) are the amplitudes of transit and reflection during its transition from the first (second) semi-infinite medium to the semi-infinite medium with zero potential. From (7) it follows that $t_{0,n}$, $r_{1,0}$ ($t_{0,2}$, $r_{0,2}$) are electron transit and reflection amplitudes for a semi-infinite medium with potential equal to that of $n$-th barrier, which is the left boundary of zero potential semi-infinite medium at the point $x_n - d_n/2$. $t_n$ and $r_n$ are electron scattering amplitudes on $n$-th rectangular potential of layer medium in case when potential is zero in all points both on left and right sides.

Below it will be shown that through denotes (5) – (8) the coefficients of solution (4) are expressed unambiguously. In accordance with the main result of transfer matrices, based on the linearity of Schroedinger equation (1), there is a linear relationship between the coefficients of solution (4) corresponding to two different space areas. Thus, following relationships exist between solution coefficients $a_n, b_n$ corresponding to zero potential areas, and the coefficient $R_{1,2}^N$ (space area occupied by the first semi-infinite medium):

$$a_n = \frac{k_{1,0}}{k_0} \left[ \left( \frac{1}{t_{1,0}^*} \frac{1}{T_{n-1}^*} + \frac{r_{1,0}}{t_{1,0}} \frac{R_{n-1}^*}{T_{n-1}^*} \right) - \left( \frac{r_{1,0}^*}{t_{1,0}^*} \frac{1}{T_{n-1}^*} + \frac{1}{t_{1,0}} \frac{R_{n-1}^*}{T_{n-1}^*} \right) R_{1,2}^N \right], \qquad (9)$$

$$b_n = \frac{k_{1,0}}{k_0} \left[ -\left( \frac{1}{t_{1,0}^*} \frac{R_{n-1}}{T_{n-1}} + \frac{r_{1,0}}{t_{1,0}} \frac{1}{T_{n-1}} \right) + \left( \frac{r_{1,0}^*}{t_{1,0}^*} \frac{R_{n-1}}{T_{n-1}} + \frac{1}{t_{1,0}} \frac{1}{T_{n-1}} \right) R_{1,2}^N \right], \qquad (10)$$

where $T_n, R_n$ ($T_0 = 1, R_0 = 0$) are electron transit and reflection amplitudes for the first $n$ rectangular potential of the lattice structure in case when potentials on left and right sides are zero:

$$\begin{pmatrix} 1/T_n^* & -R_n^*/T_n^* \\ -R_n/T_n & 1/T_n \end{pmatrix} = \prod_{l=n}^{1} \begin{pmatrix} 1/t_l^* & -r_l^*/t_l^* \\ -r_l/t_l & 1/t_l \end{pmatrix}. \qquad (11)$$

From (9) and (10) one can see that coefficients $a_n, b_n$ are defined unambiguously, if the electron reflection amplitude $R_{1,2}^N$, as well as electron scattering amplitudes for a layered structure composed only from one, two, etc. rectangular potentials $T_n, R_n$ ($n = 1, 2, ..., N$) are known.

Coefficients $c_n, d_n$ $(n=1,2,...,N)$ that correspond to space area occupied by n-th rectangular potential can be expressed through $a_n, b_n$ by following formulae:

$$c_n = \frac{k_0}{k_n}\left[\frac{1}{t^*_{0,n}}a_n - \frac{r^*_{0,n}}{t^*_{0,n}}b_n\right], \qquad (12)$$

$$d_n = \frac{k_0}{k_n}\left[-\frac{r_{0,n}}{t_{0,n}}a_n + \frac{1}{t_{0,n}}b_n\right]. \qquad (13)$$

From (12), (13) and (10), (11) it immediately follows that coefficients $c_n, d_n$, as well as $a_n, b_n$ are expressed through $R^N_{1,2}$ and $T_n, R_n$ $(n=1,2,...,N)$. Hence, we can conclude that the coefficients of solution (4) $c_n, d_n$ $(n=1,2,...,N)$, $a_n, b_n$ $(n=1,2,...,N+1)$ corresponding to the area between semi-infinite media $(0 < x < d)$ are expressed through $R^N_{1,2}$, $T_n, R_n$, and the whole problem reduces to two others: determination of transition $R^N_{1,2}$ and reflection $T^N_{1,2}$ amplitudes for the entire potential (2) and determination of electron scattering amplitudes $T_n, R_n$ for a lattice structure with lacking last N - n potentials.

The problem of determination of electron scattering amplitudes in a general form for an arbitrary potential within a finite interval bound from both sides by two different semi-infinite media was considered in [9], where algebraic relationships between the sought electron scattering amplitudes and the amplitude for the same potential for the case when left and right potentials are zero. In accordance with aforementioned denotes, these algebraic relationships can be presented in the following form:

$$\frac{1}{T^N_{1,2}} = \frac{1}{t_{1,0}}\frac{1}{t_{0,2}}\frac{1}{T_N} + \frac{r_{0,2}}{t_{0,2}}\frac{r^*_{1,0}}{t^*_{1,0}}\frac{1}{T^*_N} + \frac{r^*_{1,0}}{t^*_{1,0}}\frac{1}{t_{0,2}}\frac{R_N}{T_N} + \frac{r_{0,2}}{t_{0,2}}\frac{1}{t_{1,0}}\frac{R^*_N}{T^*_N}, \qquad (14)$$

$$\frac{R^N_{1,2}}{T^N_{1,2}} = \frac{1}{t^*_{1,0}}\frac{1}{t_{0,2}}\frac{R_N}{T_N} + \frac{r_{0,2}}{t_{0,2}}\frac{r_{1,0}}{t_{1,0}}\frac{R^*_N}{T^*_N} + \frac{r_{1,0}}{t_{1,0}}\frac{1}{t_{0,2}}\frac{1}{T_N} + \frac{r_{0,2}}{t_{0,2}}\frac{1}{t^*_{1,0}}\frac{1}{T^*_N}, \qquad (15)$$

where, according to (11), $T_n, R_n$ are scattering amplitudes for the entire lattice structure in case when potential is zero in all points left and right from the structure, i.e. $V_1 = V_2 = 0$.

From (14) it follows that the problem of definition of amplitudes $R_{1,2}^N$ and $T_{1,2}^N$ reduces to determination of $T_n, R_n$ at $n = N$. Hence, we have shown that the problem of electron wave function determination for an irregular lattice structure (2) in general reduces to determination of 2N quantities $T_n, R_n$ $(n = 1, 2, ..., N)$. The later problem was considered in [20, 21], where in particular it has been shown that the quantities under consideration are interdependent, so that following recurrent equations can be written assuming that n is a discrete variable:

$$\frac{1}{T_n} = \frac{r_n}{t_n} \frac{R_{n-1}^*}{T_{n-1}^*} + \frac{1}{t_n} \frac{1}{T_{n-1}}, \quad (16)$$

$$\frac{R_n^*}{T_n^*} = \frac{r_n^*}{t_n^*} \frac{1}{T_{n-1}} + \frac{1}{t_n^*} \frac{R_{n-1}^*}{T_{n-1}^*} \quad (17)$$

with initial conditions $T_0 = 1$, $R_0 = 0$. Note that for quantities $1/T_n$, $R_n^*/T_n^*$ (16) and (17) are linear sets of two difference equations with coefficients that contain scattering parameters of only one rectangular potential (8).

## 3. Electron Wave Function for Ideal Lattice

As shown above, the problem of electron wave function determination in a layered structure field (2) in general reduces to solution of equation set (16), (17). The latter is a difficult mathematical problem that can be solved only in some specific cases. For example, in case of equidistant and identical rectangular potentials of the lattice ($d_1 = d_2 = ... = d_N$, $U_1 = U_2 = ... = U_N$ and $x_n = x_1 + (n-1)a$, $a$ – lattice period) for $1/T_n$ and $R_n/T_n$ one has [3, 22]:

$$\frac{1}{T_n} = \exp\{ik_0 na\} \left\{ \cos n\boldsymbol{b} + i \operatorname{Im}(t_1^{-1} \exp\{-ik_0 a\}) \frac{\sin n\boldsymbol{b}}{\sin \boldsymbol{b}} \right\}, \quad (18)$$

$$\frac{R_n}{T_n} = \exp\{ik_0 (n-1)a\} \frac{r_1}{t_1} \frac{\sin n\boldsymbol{b}}{\sin \boldsymbol{b}}, \quad (19)$$

where $\cos \boldsymbol{b} = \operatorname{Re}(\exp\{-ik_0 a\}/t_1)$, which, according to (8), is

$$\cos \boldsymbol{b} = \cos k_0 (a-d) \cos kd - \frac{k^2 + k_0^2}{2k_0 k} \sin k_0 (a-d) \sin kd, \quad (20)$$

Here $k^2 = (E-U)$, $U$ - barrier potential, $d$ - barrier thickness. From (18), (19) one can see that at large values of $N$ parameter $\boldsymbol{b}$ defines the character of electron scattering on a periodic structure depending on electron energy. Thus, if $\boldsymbol{b}$ is a real quantity, then the dependence of transition coefficient on the number of system's rectangular potentials has an oscillating character, so that an energy zone of electron transition exists. Imaginary $\boldsymbol{b}$ corresponds to reflection zone, and in this case the increase of $N$ results in a complete reflection of electron. Note that expression (20) determines the spectrum of forbidden and permitted bands of electron energies for a periodical structure infinite on both sides [3].

Fig. 1 presents the graphs of electron's wave function square $|\psi|^2$ for a layered structure composed of two, four, six and eight layers (Figs. a), b), c), d) respectively) for energy $ed^2 = 4.6$, which corresponds to forbidden band , when electron's potential energy in the first and second semi-infinite media are zero: $V_1 d^2 = V_2 d^2 = 0$. From these Figs. one can see that if the number of system's potentials increases, the wave function is oscillating and tends to zero, and the envelope of maximums has an exponential character. From Figs. 2c and 2d it follows that the wave function disappears already at six barriers. It is important that the number of barriers required for disappearance of wave function strongly depends on the energy of incident electron in the reflection band. Thus, at energy value $ed^2 = 3$ the wave function disappears at four barriers. One can see that both in reflection and transition bands the square of wave function's modulus has oscillating character. However, here the envelope is periodical, versus the reflection band. For being more illustrative, on Fig. 3 we present the graphs of $|\psi|^2$ for energy values $ed^2 = 4.9$ (a) and $ed^2 = 2$ (b) at $Ud^2 = 3$, $a/d = 2$ for a layered structure with 100 layers. It can be seen that the period of $|\psi|^2$ modulation increases along with the increase of electron energy.

### Electron Wave Function for Certain Non-Ordered Structures

For a more complete illustration of possibilities of the proposed method, in this part we present the graphs of $|\psi|^2$ for certain structures, in which in various but definite

ways the ideality of layered structure is distorted, while the existence of the first and second semi-infinite media is taken into account. Fig. 4 illustrates the results of relevant calculations for four different non-ordered chains composed of $N = 8$ barriers, for which the equality of potential values, their width and distance between them are distorted

$$U_1 \neq U_2 \neq ... \neq U_N, \quad d_1 \neq d_2 \neq ... \neq d_N, \quad \Delta_1 \neq \Delta_2 \neq ... \neq \Delta_{N-1}, \quad \Delta_n = x_{n+1} - x_n - (d_{n+1} - d_n)/2$$

The distance $L$ between semi-infinite media and location of layered structure between them are chosen arbitrarily ($L \geq x_1 - d_1/2 + (d_1 + \cdots + d_N) + (\Delta_1 + \cdots + \Delta_{N-1})$). Fig. 4a corresponds to $|\psi|^2$ at electron energy $\varepsilon s^2 = 9$ (s has dimension of length), $V_1 s^2 = 2$, $V_2 s^2 = 1$, $(x_1 - d_1/2)/s = 0.75$, $(L - x_8 - d_8/2)/s = 0.75$ for a layered structure which parameters change in following regularities: $U_n s^2 = (4 + 0.35 \cdot n)$, $d_n/s = (1 + 0.1 \cdot n)$, $\Delta_n/s = (1 - 0.1 \cdot n)$, i.e. while passing from barrier to another the potential value and barrier width increase linearly, while the distance between barriers linearly decreases. Fig. 4b presents the $|\psi|^2$ graph at electron energy $\varepsilon s^2 = 3.5$, $V_1 s^2 = 2$, $V_2 s^2 = 1$, $(x_1 - d_1/2)/s = 0.75$, $(L - x_8 - d_8/2)/s = 0.75$ for a layered structure which parameters change in following regularities: $U_n s^2 = 0.05 \cdot n^2$, $d_n/s = (1 + 0.1 \cdot n^2)$, $\Delta_n/s = (1 + 0.1 \cdot n)$, i.e. the potential value and barrier width increase squared, while the distance between barriers increases linearly. $|\psi|^2$ graph on Fig. 4c corresponds to following values of problem parameters: $\varepsilon s^2 = 1.3$, $V_1 s^2 = 0.5$, $V_2 s^2 = 0.75$, $(x_1 - d_1/2)/s = 1$, $(L - x_8 - d_8/2)/s = 1$, $U_n s^2 = 0.035 \cdot n(m - n + 1)$ ($m = 8$), $d_n/s = 0.2n + 0.1n^2$, $\Delta_n/s = 0.1 \cdot n^2$. The $|\psi|^2$ graph on Fig. 4d corresponds to electron energy $\varepsilon s^2 = 5$ for a lattice composed of equidistant barriers of equal width with periodically modulated potentials. The structure parameters are chosen as follows: $V_1 s^2 = 0.5$, $V_2 s^2 = 0.75$, $(x_1 - d_1/2)/s = 1$, $(L - x_8 - d_8/2)/s = 1$, $U_n s^2 = 4 \cdot \sin^2 n$ ($m = 8$), $d_n/s = 1$, $\Delta_n/s = 1$. It should be noted that for the first three cases on Fig. 4, unlike the last case, the areas of electron transition and reflection do not sequent one another for a few times, as it was in case of periodic potential. One can state that, starting from zero, there is a range of energy values within which the probability of electron transition is practically zero everywhere, except a few points where it does not exceed 3 – 5%. Above this energy range electron starts to pass through the structure, and the turn of reflection band into transition one is not sharp, since they are separated by an intermediate zone. It is also noteworthy, that the increase of scattering barriers leads to expansion of reflection band. This is due to well known phenomenon of wave localization in a 1D non-ordered structure, and the larger is electron's energy, the is $|\psi|^2$ modulation period.

**Conclusion**

In the present paper we have shown that the problem of determination of the wave function for electron's infinite motion in the field of 1D irregular structure in general can be reduced to solution of a certain set of two difference equations. The proposed approach has been applied to two cases: ideal and non-ordered lattices.

The consideration of the first case revealed that the square of wave function's modulus for energies corresponding to reflection band, oscillates and tends to zero when the number if system's potentials increases, and the envelope of maximums decreases exponentially. Important is that the number of barriers required for annulling of the wave function, strongly depends on the energy of incident electron in the band. The wave function for transition band, similar to reflection band, again has an oscillating character, however, here the envelope of modulus' square has a periodical character. The more is electron energy within the permitted band, the more is $|\psi|^2$ modulation period.

To illustrate the opportunities provided by the proposed approach, we presented $|\psi|^2$ graphs for certain non-ordered lattices in which by some ways the ideality of layered structure is distorted, while the existence of first and second semi-infinite media is taken into account. The consideration revealed that for some structures a range of energy values exist where the areas of electron transition and reflection do not sequent one another for a few times, as it was in case of periodic potential. One can state that, starting from zero, there is a range of energy values within which the probability of electron transition is practically zero everywhere, except a few points where it does not exceed 3 – 5%. Above this energy range electron starts to pass through the structure, and the turn of reflection band into transition one is not sharp, as it is in case of periodic structure, since they are separated by an intermediate zone, and the increase of the number of scattering barriers leads to an expansion of reflection band.

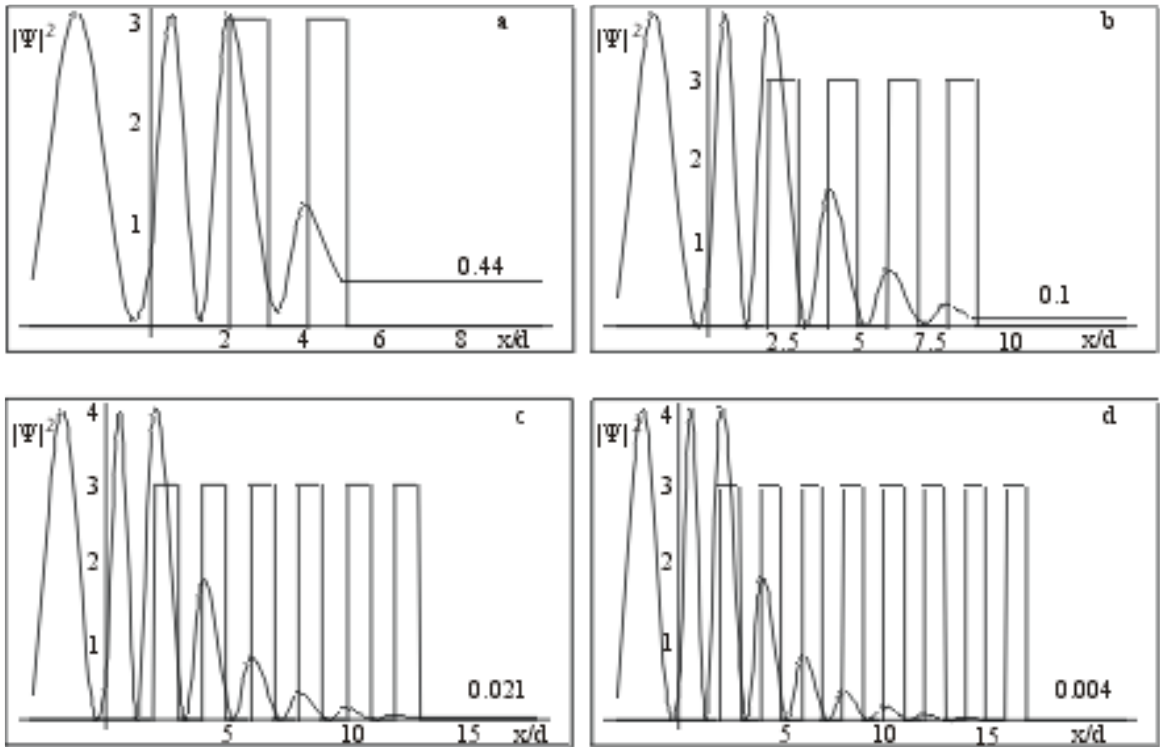

Fig. 1. Square of electron wave function's modulus for electron energy $ed^2 = 4.6$ corresponding to the second forbidden band at $Ud^2 = 3$, $a/d = 2$ for a layered structure with different numbers of barriers. Corresponding transition coefficients are shown in the right bottom corner.

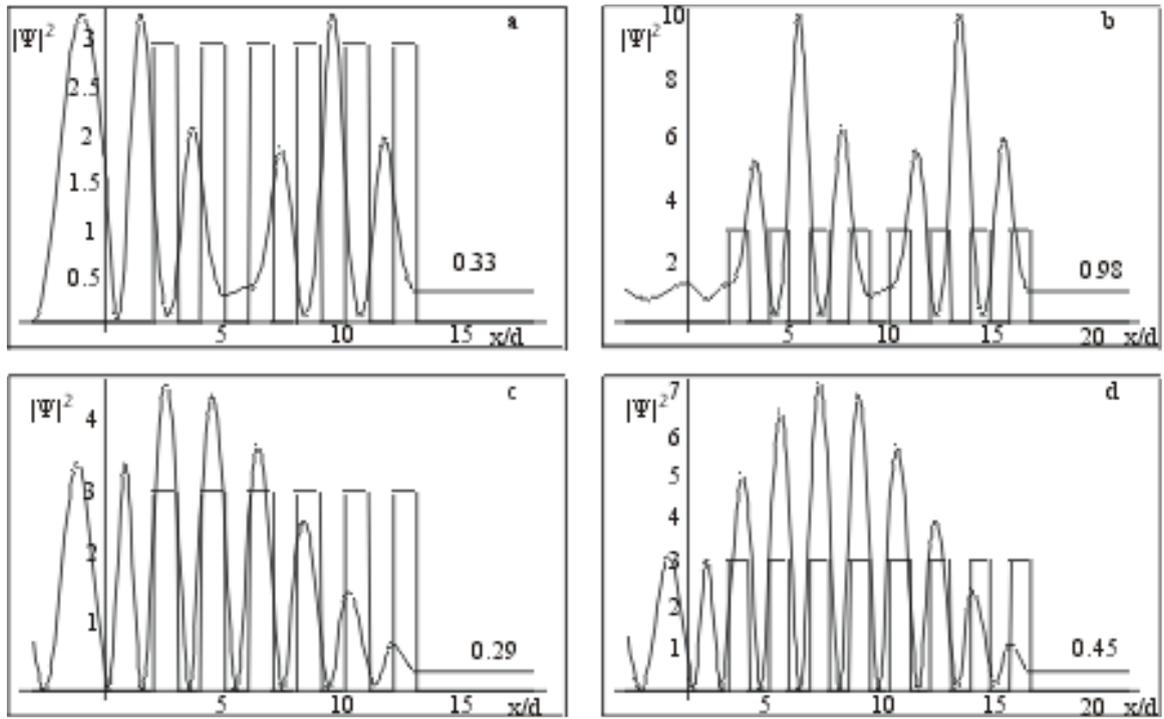

Fig. 2. Square of electron wave function's modulus for electron energies $ed^2 = 2.5$ and $ed^2 = 5$ corresponding to first and second forbidden bands (Figs. a), b), c), d) respectively), at $Ud^2 = 3$, $a/d = 2$ for a layered structure with different numbers of barriers. Corresponding transition coefficients are shown in the right bottom corner.

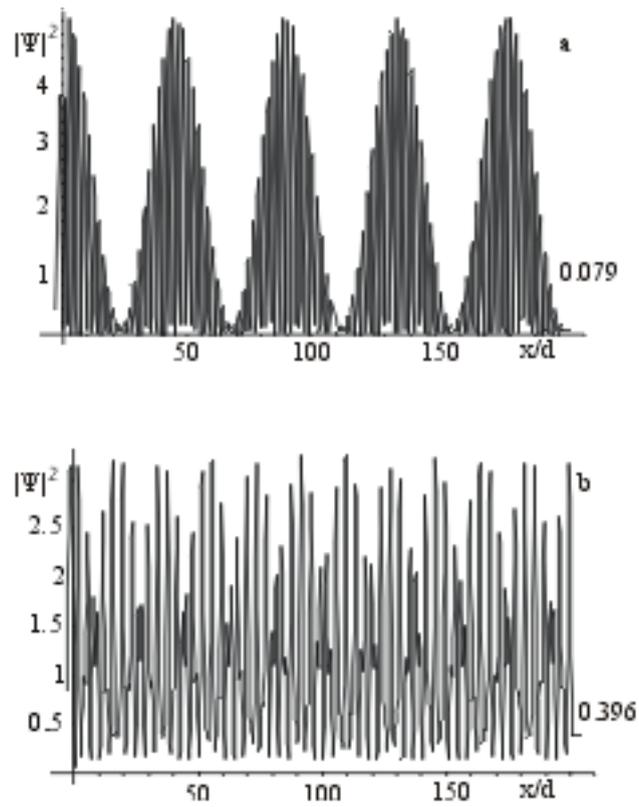

Fig. 3. Square of electron wave function's modulus for electron energies $\varepsilon d^2 = 4.9$ (a) and $\varepsilon d^2 = 2$ (b) at $Ud^2 = 3$, $a/d = 2$ for a layered structure with 100 barriers. Corresponding transition coefficients are shown in the right bottom corner.

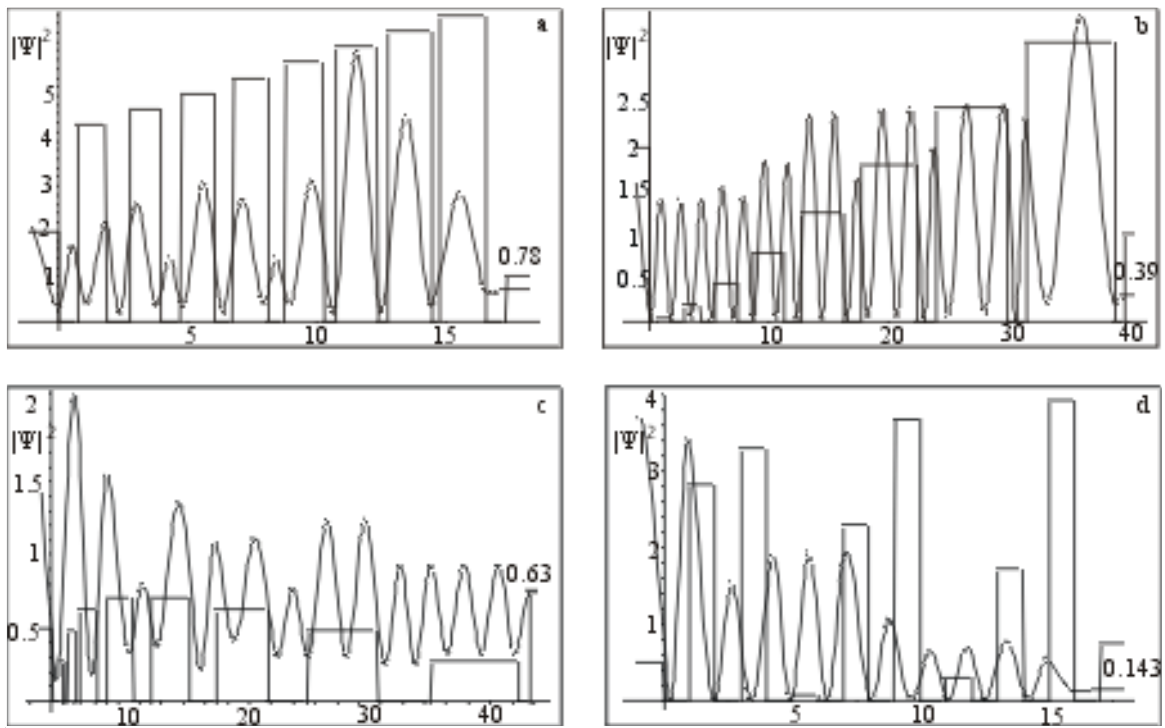

Fig. 4. Square of electron wave function's modulus for certain non-ordered structures.